\documentclass[%
 reprint,
 amsmath,amssymb,
 aps,
 superscriptaddress,
]{revtex4-1}

\usepackage{graphicx}
\usepackage{dcolumn}
\usepackage{bm}
\usepackage{hyperref}

\begin{document}	
	\title{Large spin gaps in half metals MN$_4$ (M=Mn, Fe, Co) with N$_2$ dimers}
	
	\author{Jun Deng}
	\affiliation{Beijing National Laboratory for Condensed Matter Physics, Institute of Physics, Chinese Academy of Sciences, Beijing 100190, China}
	\affiliation{University of Chinese Academy of Sciences, Beijing 100049, China}
	\author{Ning Liu}
	\affiliation{Beijing National Laboratory for Condensed Matter Physics, Institute of Physics, Chinese Academy of Sciences, Beijing 100190, China}
	\affiliation{University of Chinese Academy of Sciences, Beijing 100049, China}
	\author{Jiangang Guo}
	\email{jgguo@iphy.ac.cn}
	\affiliation{Beijing National Laboratory for Condensed Matter Physics, Institute of Physics, Chinese Academy of Sciences, Beijing 100190, China}
	\affiliation{Songshan Lake Materials Laboratory , Dongguan, Guangdong 523808, China}
	\author{Xiaolong Chen}
	\email{xlchen@iphy.ac.cn}
	\affiliation{Beijing National Laboratory for Condensed Matter Physics, Institute of Physics, Chinese Academy of Sciences, Beijing 100190, China}
	\affiliation{University of Chinese Academy of Sciences, Beijing 100049, China}
	\affiliation{Songshan Lake Materials Laboratory , Dongguan, Guangdong 523808, China}
	\date{\today}

	\begin{abstract}
		We predict that cubic MN$_4$ (M=Mn, Fe, Co) are all half metals with the largest spin gap up to $\sim$ 5 eV. They possess robust ferromagnetic ground states with the highest Curie temperature up to $\sim$ 10$^3$ K. Our calculations indicate these compounds are energetically favored, dynamically and mechanically stable. It is proposed that self-doping of these 3\emph{d} transition metals occurs in MN$_4$ due to the reduction in electronegativity of N$_2$ dimers. This model can well explain the calculated integer magnetic moments, large spin gaps of MN$_4$ and semiconducting behavior for NiN$_4$ as well. Our results highlight the difference in electronegativity between transition metal ions and non-metal entities in forming half metals and the role of N$_2$ dimer in enlarging the spin gaps for nitride half metals. 
		\begin{description}
			\item[PACS numbers]
			blank
		\end{description}
	\end{abstract}
	\pacs{blank}
	\maketitle
\section{Introduction}
	Half metals are a class of material that behave as metals by electrons of one spin orientation and as semiconductors by electrons of the other spin orientation \cite{RN20}. They are promising candidates for spintronics applications from magnetic tunneling junctions to giant magnetoresistance devices and injecting spin-polarized currents into semiconductors \cite{RN28,RN32,RN35}. For practical applications, ideal half metals should be magnets having high Curie temperatures (\emph{T}$_{c}$) and large enough half-metallic gaps. Previous studies identified a number of half metals with varying \emph{T}$_c$ ($\sim$1100K) and gap (0.5 eV $\sim$ 2 eV), including full heusler Co$_2$FeSi \cite{RN24}, half heusler NiMnSb \cite{RN30}, oxides CrO$_2$ \cite{RN29} and Fe$_3$O$_4$ \cite{RN25}, perovskites Sr$_2$FeMoO$_6$ \cite{RN21} and so on, which have been intensively investigated over the last decades. The search and prediction of new half metal with better performance, however, are still tough tasks. 
	
	So far, all known half-metallic candidates lie in a limited structural types and chemical species. They can be basically categorized into two groups: one being high-valence transition metal oxides, the other low-valence transition metal silicides or antimonides. Both groups of half metals contain one or two transition metals centered either in tetrahedra or octahedra in their structures. Ferro- and ferrimagnetism are realized through double-exchange or super-exchange interactions of spins affected by the effects of Hund's rule, crystal field and orbital hybridization. Two channels of band structure, metallic and non-metallic characteristics, form in terms of spin direction as the most striking feature for half metals. In the non-metallic channel, the valence bands (VB) are always composed of half-filled five \emph{d} or three \emph{p} orbital electrons, resulting in full occupation of the bands in one spin direction. These \emph{d} electrons come from a single magnetic Fe ion in Sr$_2$MoFeO$_6$ \cite{RN26} or two magnetic ions as Ni and Mn in Heusler NiMnSb \cite{RN30}. The \emph{p} orbitals usually come from oxygen in the case for CrO$_2$ \cite{RN78}. Their conduction bands (CB) are, therefore, empty. In the metallic channel, the energy bands composed of hybrid orbitals of \emph{d} metals or/and non-metal elements cross the Fermi energy.
	
	From an electronegative point of view, these half-metals can be regarded as formation from high valence  transition metal ions and a non-metallic element with large electronegativity, or from low-valcence transition metal ions and  non-metallic elements with relative low electronegativity. High-valence transition metals in oxides are difficult further to transfer electrons to oxygen or via versa because of metals' high ionization energy for remaining electrons and large electronegativity for oxygen. So do for the second group half metals with low-valence transition metals where the hopping of electrons are forbidden due to the weak bonding of transition metals and low-electronegative Si and Sb.  If a transition metal has a proper number of electrons in its \emph{d} subshell to guarantee a full occupation of VB half-metallicity will ensue. A N$_2$ dimer has a lower electronegativity in comparison with atomic N as it meets the octet electron rule. Moreover, the $\pi$ and $\pi^*$ orbitals in the dimer may serve as the VB and CB with a large gap in the non-metallic channel, helpful in widening the spin gaps for half metals. This proposition deserves a test for there is no known half metals with a dimer as its non-metallic component up to now.

\begin{table*}
	\caption{\label{table1}Computed cohesive energies for MN$_4$ and known M-N compounds (M=Fe, Co, Mn).}
	\begin{ruledtabular}
		\begin{tabular}{lclclc}
			Compounds& $\emph{E}$$_c$ (eV/atom)& Compounds& $\emph{E}$$_c$ (eV/atom)& Compounds& $\emph{E}$$_c$ (eV/atom)\\	
			\hline
			FeN$_4$($\emph{Fd$\bar{3}$m}$)& -5.079& CoN$_4$($\emph{Fd$\bar{3}$m}$)& -5.182& MnN$_4$($\emph{Fd$\bar{3}$m}$)& -4.867\\
			FeN($\emph{F$\bar{4}$3m}$) \cite{RN79}&-5.188& CoN($\emph{F$\bar{4}$3m}$) \cite{RN89}& -5.297&MnN($\emph{I4/mmm}$) \cite{RN92}&-4.737\\
			Fe$_2$N($\emph{Pbcn}$)\cite{RN84}&-4.991&Co$_2$N($\emph{Pnnm}$) \cite{RN90}&-5.393&Mn$_3$N$_2$($\emph{I4/mmm}$) \cite{RN93}&-4.703\\
			Fe$_2$N($\emph{P$\bar{3}$m1}$) \cite{RN94}&-4.930&Co$_3$N($\emph{P6$_3$22}$) \cite{RN91}&-5.416&Mn$_4$N($\emph{Pm$\bar{3}$m}$) \cite{RN83} &-4.009\\
			Fe$_2$N($\emph{P312}$) \cite{RN85}&-5.050&Co$_4$N($\emph{Pm$\bar{3}$m}$) \cite{RN91}&-5.423&\\
			Fe$_2$N($\emph{P$\bar{3}$1m}$) \cite{RN95}&-5.050\\
			Fe$_3$N($\emph{P6$_3$22}$) \cite{RN86}&-5.063\\
			Fe$_3$N($\emph{P312}$) \cite{RN85}&-5.063\\
			Fe$_4$N($\emph{Pm$\bar{3}$m}$) \cite{RN87}&-5.015\\
			Fe$_8$N($\emph{I4/mmm}$) \cite{RN88} &-4.981
		\end{tabular}
	\end{ruledtabular}
\end{table*}
	
	Recently, a new cubic compound SiN$_4$ was predicted to be both thermodynamically and lattice-dynamically stable. Its structure consists of SiN$_4$ tetrahedra connected by N$_2$ dimers \cite{RN51}.  In this study, by first principles calculations, we examined the above-mentioned proposition and half-metallicity for MN$_4$ (M=Mn, Fe, Co, and Ni). Our results indicate that they are all half metals with largest spin gap up to $\sim$ 5 eV and highest \emph{T}$_c$ $\sim$ 10$^3$ K except for Ni. We propose that the transition metals are self-doped with their 4\emph{s} electrons due to the relatively low electronegativity of the N$_2$ dimer and the ferromagnetism arises by their spin-parallel 3\emph{d} electrons in the metallic channel. In the non-metallic channel, the large spin gaps can be attributed to the big energy difference between the bonding and antibonding states of N$_2$ dimers.

\section{Computational methods}
	Geometry optimizations, phonon spectra, band structures were performed using the density functional theory with the generalized gradient approximation (GGA) in the form of the Perdew-Burke-Ernzerhof (PBE) \cite{RN17} exchange-correlation potential implemented in CASTEP \cite{RN19}. Mechanical properties, orbital resolved band structures were calculated by using Vienna \emph{ab initio} simulation Package (VASP) \cite{RN18}. The Hubbard repulsion term \emph{U} was introduced to account for the correlated effects of 3\emph{d} electrons. In calculation of band gaps, the HSE06 hybrid functional was adopted to enhance the accuracy. The detailed calculation methods can be found at the Supplemental Material \cite{SI}. 

\section{Results and discussion}
\begin{figure}
	\includegraphics{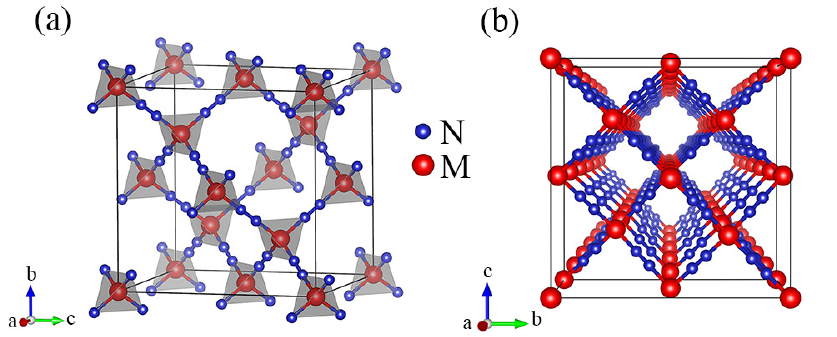}%
	\caption{\label{fig1}(a) Crystal structure of MN$_4$ (M is Mn, Fe, Co), space group $\emph{Fd$\bar{3}$m}$ ($\emph{No.227}$), which can be regarded as formation from the replacement of the C atoms in diamond by MN$_4$ tetrahedron adapted from $\emph{Ref.}$ \cite{RN51}. (b) Perspective view from the $\lbrack$100$\rbrack$ direction of MN$_4$.}
\end{figure}

	As described in $\emph{Ref.}$ \cite{RN51}, we build MN$_4$ with a diamond-structure (space group $\emph{Fd$\bar{3}$m}$) shown in Fig. \ref{fig1}(a), where M is Mn, Fe and Co. In each unit cell, there are eight MN$_4$ tetrahedral units that are connected by N-N bonds. The M ions occupy the Wyckoff 8(a) site, N 32(e) site. After optimization, the M-N and N-N bondlengths are slightly shortened compared with the known compounds \cite{RN79,RN47,RN40} and diazenides \cite{RN76,RN77}. Their structural parameters are summarized in Table S1 \cite{SI}.

\begin{figure}
	\includegraphics{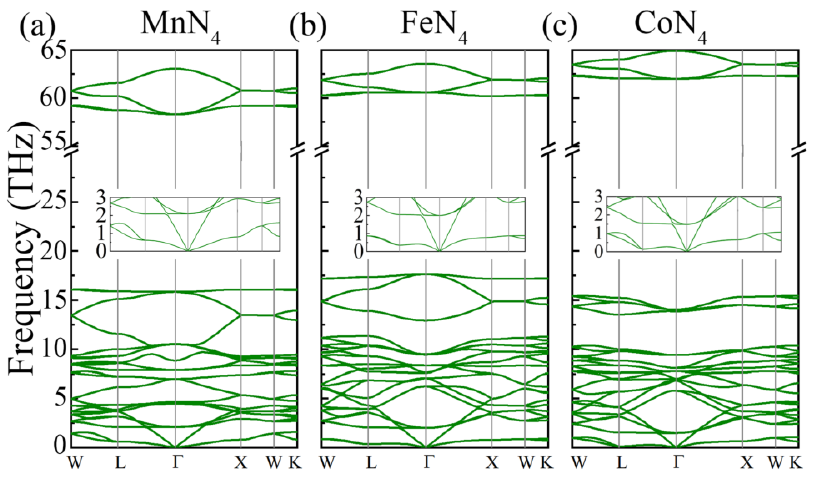}%
	\caption{\label{fig2}Phonon spectra for (a) MnN$_4$, (b) FeN$_4$ and (c) CoN$_4$. The insets are zoom-in images of low frequencies of phonon spectra.}
\end{figure}

\begin{figure}
	\includegraphics{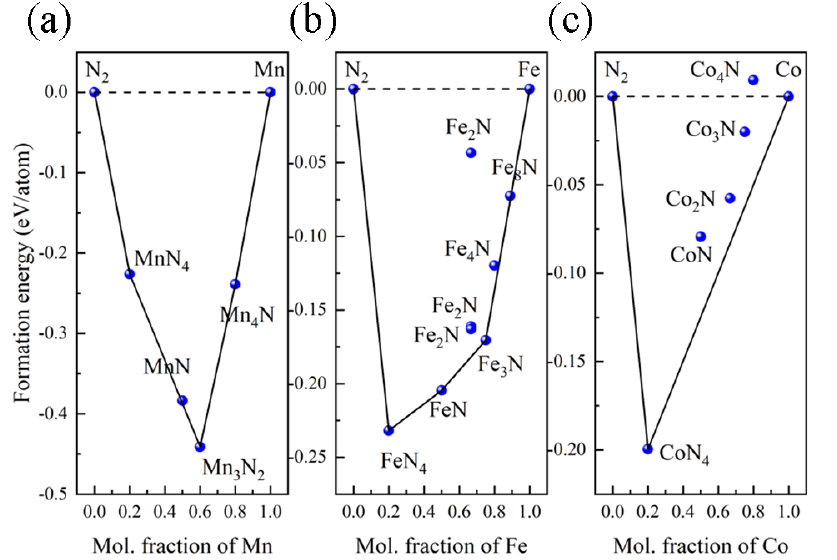}%
	\caption{\label{fig3}Formation energies of the existing (a) Mn-N, (b) Fe-N, (c) Co-N compounds and MN$_4$ (M=Mn, Fe, Co) respect to decomposition into their elemental states. The convex hulls are shown by solid line. The data points on the solid line means the structure is stable.}
\end{figure}

	 The stability of these compounds were examined by phonon spectra, in which no negative frequencies appear, suggesting these compounds are lattice-dynamically stable, see Fig. \ref{fig2}(a), \ref{fig2}(b) and \ref{fig2}(c). Then we evaluate the cohesive energies $\emph{E}$$_c$ for MnN$_4$, FeN$_4$, and CoN$_4$ along with other Mn, Fe, and Co nitrides for comparison. $\emph{E}$$_c$ is defined as $E_c=(E_{M_xN_y}-xE_M-yE_N)/(x+y)$, where $E_M$, $E_N$, and $E_{M_xN_y}$ are the total energy of a single M atom, a single N atom, and $M_xN_y$ compound. The results are listed in Table \ref{table1}. MnN$_4$, FeN$_4$, CoN$_4$ are either more negative or close to $\emph{E}$$_c$ of the known binary compounds,  indicating they are energetically stable in terms of cohesive energy. On the other hand, a convex hull analysis based on formation energy is performed. The formation energy is defined as $E_f=(E_{M_xN_y}-x\mu_M-y\mu_N)/(x+y)$,where $\mu_M$, $\mu_N$, and $E_{M_xN_y}$ are the total energy of stable bulk crystal A, crystal B, and $A_xB_y$ compound. Fig. \ref{fig3} along with Table S2 \cite{SI} shows that these MN$_4$ are also stable. Furthermore, for a stable cubic structure, the elastic matrix should satisfy the Born stability criteria \cite{RN31}: $\emph{C}$$_{11}$-$\emph{C}$$_{12}$$>$0, $\emph{C}$$_{11}$+2$\emph{C}$$_{12}$$>$0, $\emph{C}$$_{44}$$>$0. The calculated $\emph{C}$$_{11}$,$\emph{C}$$_{12}$ and $\emph{C}$$_{44}$, summarized in Table S3 \cite{SI}, meet well these criteria. The structural stability of FeN$_4$ at elevated temperatures are examined through molecular dynamics calculations, as shown in Fig. S1 \cite{SI}. The results indicate that the structure of FeN$_4$ is well preserved up to 900 K.

\begin{table*}
	\caption{\label{table2}Half-metallic gaps (PBE) and $\emph{T$_c$}$s for MnN$_4$, FeN$_4$, CoN$_4$ and known compounds.}
	\begin{ruledtabular}
		\begin{tabular}{ccccccccc}
			Compounds&MnN$_4$&FeN$_4$&CoN$_4$&CrO$_2$&Sr$_2$FeMoO$_6$&NiMnSb&Fe$_3$O$_4$&Co$_2$FeSi\\
			\hline
			$\emph{T$_c$ }$ (K)&1.90$\times$10$^3$&2.70$\times$10$^3$&8.18$\times$10$^2$&386 \cite{RN96}&419 \cite{RN97}&730 \cite{RN98}&851 \cite{RN99}&1100 \cite{RN100}\\
			Gap (eV)&2.36&2.65&	2.56&		$>$1.5 \cite{RN101}& 	0.8 \cite{RN102}& 	0.4 \cite{RN103}& 	0.5 \cite{RN104}&0.86 \cite{RN105}\\ 
		\end{tabular}
	\end{ruledtabular}
\end{table*}

	 The ground magnetic state of MN$_4$ is determined by calculating the total energies of ferromagnetic (FM), antiferromagnetic (AFM), and non-magnetic(NM) states. The FM state is obtained from collinear spin-polarized optimization of identical spin directions of the M atoms (see Fig. S2(a) \cite{SI}). The AFM state is just reversing the nearest M spin directions in the lattice (see Fig. S2(b) \cite{SI}) and the NM state a non-spin-polarized optimization. Whether or not considering the Hubbard \emph{U} term, the FM is found to be the ground state with local moment 3$\emph{$\mu$}$$_B$, 2$\emph{$\mu$}$$_B$, 1$\emph{$\mu$}$$_B$  per formula unit for MnN$_4$, FeN$_4$, CoN$_4$, respectively. The FM coupling strength, or the Curie temperature $\emph{T}$$_{c}$, can be estimated by the mean-field approximation (MFA) through  $\emph{k}$$_B$$\emph{T}$$_{c}$$^{MFA}$=2$\emph{$\Delta$E}$/3\emph{N} \cite{RN37}, where $\emph{k}$$_B$ is the Boltzmann constant, \emph{N} the number of magnetic atoms in the unit cell and $\emph{$\Delta$E}$ the energy difference between AFM and FM states. This formula, however, usually overestimates $\emph{T}$$_{c}$. By using empirical relationship $\emph{T}$$_{c}$/$\emph{T}$$_{c}$$^{MFA}$=0.8162 \cite{RN54}, we correct the $\emph{T}$$_{c}$s list them in Table \ref{table2}. All $\emph{T}$$_{c}$s are well above room temperature, especially for FeN$_4$ with \emph{T}$_c$=2.70$\times$10$^3$ K, much higher than that for known ferromagnets.
	
\begin{figure}
		\includegraphics{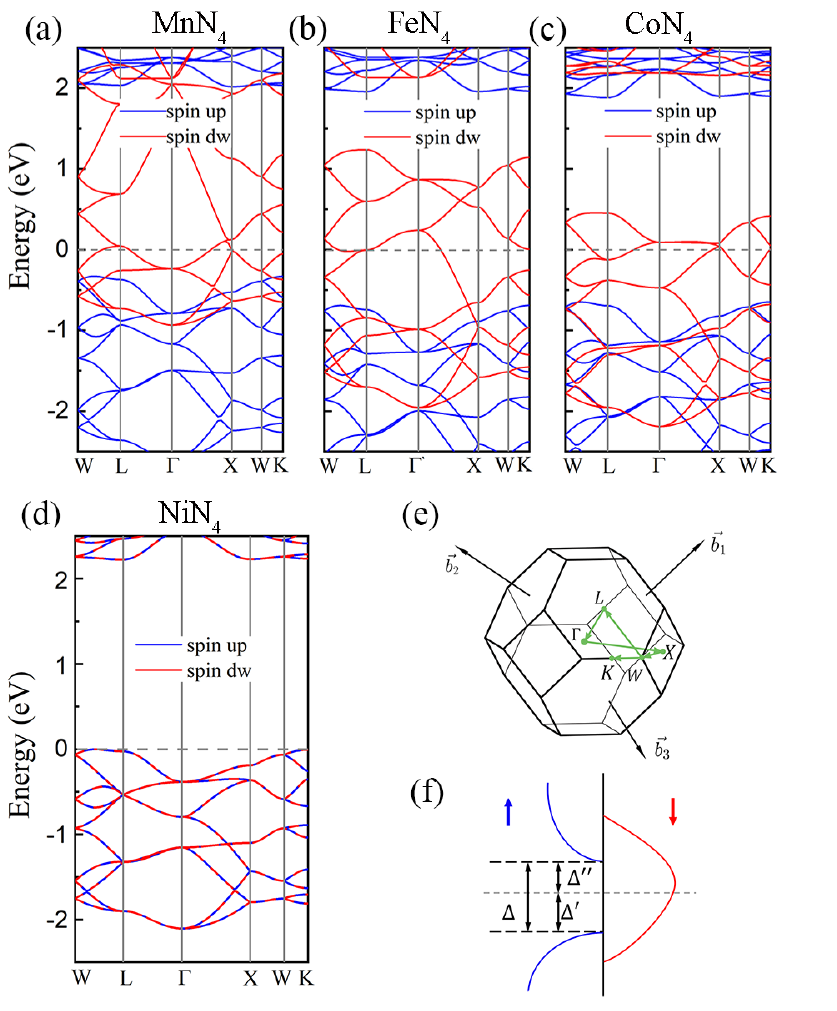}%
		\caption{\label{fig4}Spin resolved band structures for (a) MnN$_4$, (b) FeN$_4$, (c) CoN$_4$ and (d) NiN$_4$ with PBE functional. Blue lines denote the band structure of spin-up sub-band, while red spin-down (dw) sub-band. (e) First Brillouin zone. (f) Spin resolved density of state schematic, spin gaps are marked as $\Delta$, $\Delta$$^{'}$ and $\Delta$$^{''}$.}
	\end{figure}

\begin{figure}
		\includegraphics{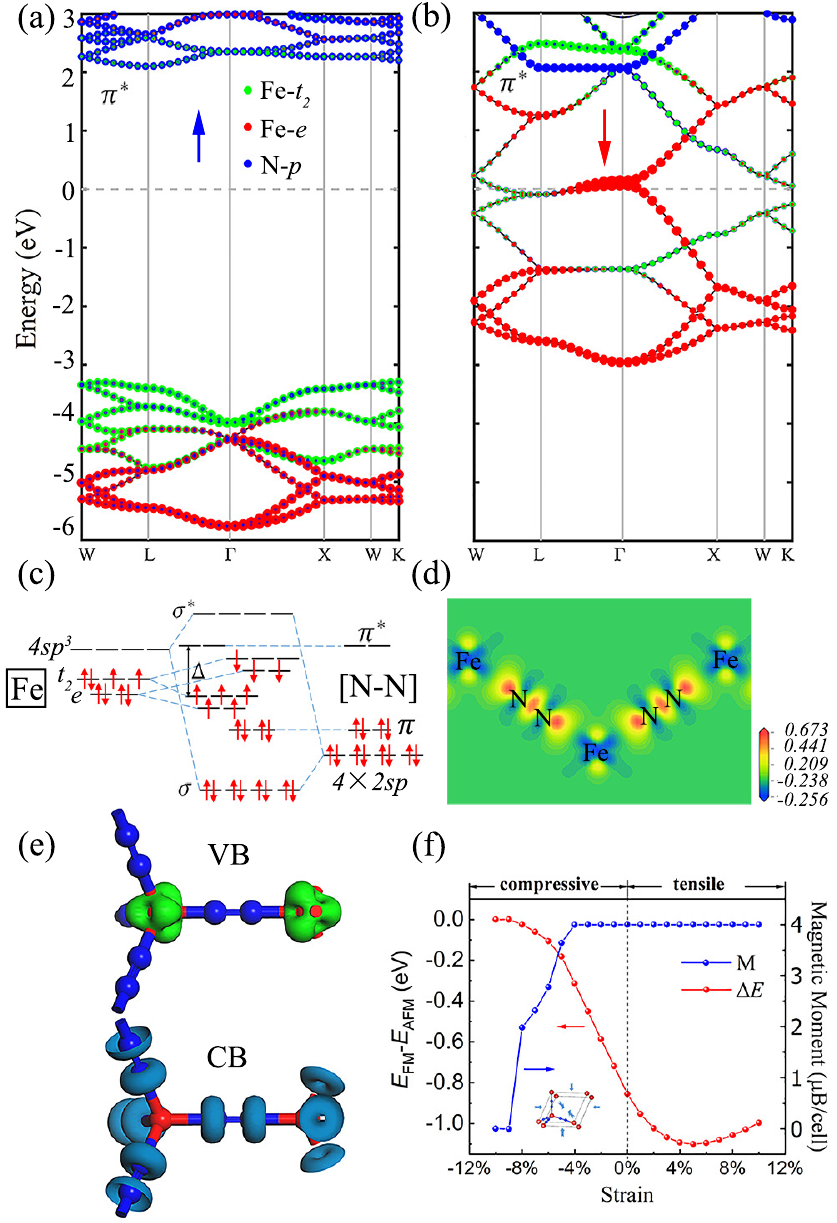}%
		\caption{\label{fig5}Orbital resolved band structure for (a) spin up and (b) spin down of FeN$_4$ calculated with HSE06 functional. Solid red and green circles represent $\emph{e}$ and $\emph{t}$$_2$ orbitals for Fe atoms, and solid blue $\emph{p}$ orbitals for N atoms. The larger the circle is, the greater an orbital contributes. (c) Schematic illustration of bonding states between an Fe atom and N$_2$ dimers. Fe 4$\emph{s}$ electrons falls to 3$\emph{d}$ orbitals and Fe-4$\emph{s}$, 4$\emph{p}$ orbitals hybrid with four N-2$\emph{sp}$ hybridization orbitals, leaving 8 electrons in its 3$\emph{d}$ orbitals. (d) Charge density difference map for FeN$_4$ along the (110) plane, obtained by $\Delta$$\rho$=$\rho$$_{FeN_4}$-$\rho$$_{Fe}$-$\rho$$_N$. (e) Electron density of the VB (upper panel) and the CB (lower panel) for the non-metallic channel. (f) The variation of energy difference between FM and AFM states (red) and magnetic moment (blue) under different strain.}
	\end{figure}

	 Fig. \ref{fig4} (a-c) show the spin resolved band structures of MN$_4$ (M=Mn, Fe, Co) calculated by PBE exchange functional, where the Fermi level is set to be zero. In the spin-down sub-band, a metallic feature shows up. While in the spin-up sub-band, a semiconducting feature with varying gap emerges. The calculated $\emph{T}$$_{c}$s and gaps of MN$_4$ and known half metals are summarized in Table \ref{table2}. Among them, FeN$_4$ owns the largest spin gap $\sim$ 2.65 eV. The effect of the on-site Hubbard $\emph{U}$ of 3$\emph{d}$ metals should be taken into account. If we adopt $\emph{U}$=3.8 eV for Fe, a modest value as in $\emph{Ref.}$ \cite{RN62}, this yields a gap $\sim$ 3.63 eV, see Fig. S3(b) \cite{SI}. To get a more accurate half-metallic gap for FeN$_4$, we performed the calculation using the HSE06 hybrid functional (see Fig. \ref{fig5}(a) and \ref{fig5}(b)) and got a spin gap $\Delta$=5.39 eV. It is noted that the energy spans, labeled $\Delta$$^{'}$ and $\Delta$$^{''}$ in Fig. \ref{fig4}(f), are 3.29 eV and 2.10 eV, respectively, which are wide enough to prevent spin-flip transitions by thermal excitations. 
	
 	Apart from MnN$_4$, FeN$_4$, CoN$_4$, extension to other isostructural 3$\emph{d}$ transition metals VN$_4$, CrN$_4$, and NiN$_4$ is tried. Again, no any negative frequencies are present in their phonon spectra for CrN$_4$ and NiN$_4$ (see Fig. S4 \cite{SI}). For VN$_4$, a small negative value $\sim$ 0.3 THz is present near point L, which can be neglected in comparison with its large positive frequencies. In contrast, considerably negative frequencies exist for ScN$_4$, TiN$_4$, CuN$_4$ and ZnN$_4$. Among the lattice-dynamically stable structures, none is magnetic and VN$_4$, CrN$_4$ are normal metals (Fig. S5 \cite{SI}). Of particular interest, NiN$_4$ is a semiconductor, see Fig. \ref{fig3}(d). Their lattice parameters and properties are summarized in Table S1 \cite{SI}.

 	Now, we try to understand the origin of the half-metallicity of MN$_4$.  Fig. \ref{fig5}(a) and \ref{fig5}(b) show the orbital resolved band structure of non-metallic and metallic channels for FeN$_4$, respectively. The 3$\emph{d}$ orbitals of Fe atoms are split into two groups, $\emph{t}$$_2$ orbitals ($\emph{d$_{xy}$}$, $\emph{d$_{yz}$}$, $\emph{d$_{xz}$}$) and $\emph{e}$ orbitals ($\emph{d$_{x^2-y^2}$}$ and $\emph{d$_{z^2}$}$), which is consistent with the situation under a tetrahedral crystal field. Apart from the $\emph{sp}$ hybrid orbitals, $\pi$ and $\pi$$^*$ orbitals are expected to form in N$_2$ dimers since electron density accumulations in between N-N dimers can be clearly seen in Fig. \ref{fig5}(d). In the non-metallic channel, the VB are composed mainly of the Fe $\emph{t}$$_2$ orbitals and the CB of the N $\pi$$^*$ orbitals. It is noted that the energy is weakly dispersed in the VB, suggesting Fe 3$\emph{d}$ are not strongly bonded with the N$_2$ dimers. A similar situation occurs in the bands just above the Fermi energy in the other channel. The $\pi$ bands (not shown) in both channels are far below the Fermi energy due to the wide energy separation of the $\pi$ and $\pi$$^*$ orbitals in the N$_2$ dimers. 
 	
 	Here we propose a model showing the bonding states between an Fe atom and four N$_2$ dimers sketched in Fig. \ref{fig5}(c). As the Fe atom is coordinated by four N$_2$ dimers, four $\sigma$ bonds might form from Fe-4$\emph{sp$^3$}$ orbitals and N-2$\emph{sp}$ orbitals. Since Fe-4$\emph{sp$^3$}$ is much higher than N-2$\emph{sp}$ in energy, these electrons in the $\sigma$ bonds can be totally contributed by the latter. A similar hybridization of Co-Si bond in Heusler Co$_2$MnSi \cite{RN23}  was observed. If this is true for the Fe-N $\sigma$ bonds, then Fe 4$\emph{s}$ electrons can only flow into its $\emph{d}$ orbitals. That is the so-called self-doping, resulting in an Fe 3\emph{d} electron configuration 
	$\uparrow$$\uparrow$$\uparrow$$\uparrow$$\uparrow$$\downarrow$$\downarrow$$\downarrow$,  in good agreement with the calculated magnetic moment 2$\emph{$\mu$}$$_B$ per FeN$_4$ formula unit. More importantly, in the non-metallic channel, the Fe 3$\emph{d}$ orbitals are fully occupied by five electrons, giving rise to a completely filled VB and an empty CB as the $\pi$$^*$ orbitals are far higher in energy. In this way a large gap appears as labeled $\Delta$ in Fig. \ref{fig5}(c). The charge density shown in Fig.  \ref{fig5}(e) confirms that the main occupancy of the Fe 3\emph{d} electrons in the VB and the empty of N-N antibonding states in the CB. On the contrary, in the metallic channel, Fe 3$\emph{d}$ orbitals are only occupied by three electrons, hence the corresponding band crosses the Fermi energy. 

	This model could theoretically explain the integer magnetic moments for MnN$_4$, CoN$_4$ and NiN$_4$. Since Mn and Co have one less and one more 3$\emph{d}$ electron compared with Fe, their moments are, therefore, 3$\emph{$\mu$}$$_B$ and 1$\emph{$\mu$}$$_B$, respectively. In both cases, their VBs are fully occupied by five 3$\emph{d}$ electrons in their non-metallic channels. For Ni, its 3$\emph{d}$ orbitals will be occupied by eight 3$\emph{d}$ and two 4$\emph{s}$ electrons, leading to full occupations in both channels. Hence, it is a semiconductor.
 
 	Considering spin interactions in a linear Fe-N$_2$-Fe way, a double-exchange like mechanism may be applicable here to establish the long-range magnetic ordering, see Fig. S6 \cite{SI}. If one spin-down electron hops from Fe 3\emph{d} orbitals to the neighboring N$_2$ dimer $\pi^*$ orbitals by thermal excitations in the metallic channel, the vacancy will be filled by one spin-down electron from another neighboring N$_2$ dimer. This is rather probable because these orbitals are overlapped to some extent, see Fig. \ref{fig5}(b). This way will help retain the spin direction and hence maintain the FM state.  
	
	The self-doping phenomena of Fe can be understood from the electronegative point of view. A N$_2$ dimer, as a chemical spices, is quite inert and low electronegative as it meets the octet electron rule. It, however, can act as a non-metallic element and form a variety of compounds with active metals. The reason is that a N$_2$ dimer has empty $\pi$$^*$ orbital, which can accommodate electrons from active metals. Such metals are usually alkali and alkaline-earth metals with high-enough energy \emph{s} electrons. Typical examples are SrN and SrN$_2$ \cite{RN48}, both containing N$_2$ dimers in their structures and electrons in the $\pi^*$ orbitals. Similarly, electron transfer to O$_2$ dimer occurs in magnetically frustrated Rb$_4$O$_6$ \cite{RN42,RN39,RN82}. Here less active Fe has a little stronger electronegativity than Sr, whose 4\emph{s}$^2$ electrons are expected not to transfer to the N$_2$ dimers' $\pi$$^*$ orbitals. Instead, they go to Fe's 3\emph{d} orbitals, that is, the self-doping of Fe. Our above calculations support that in the ground state, the $\pi$$^*$ orbitals are unoccupied. 
	
\begin{figure}
    	\includegraphics{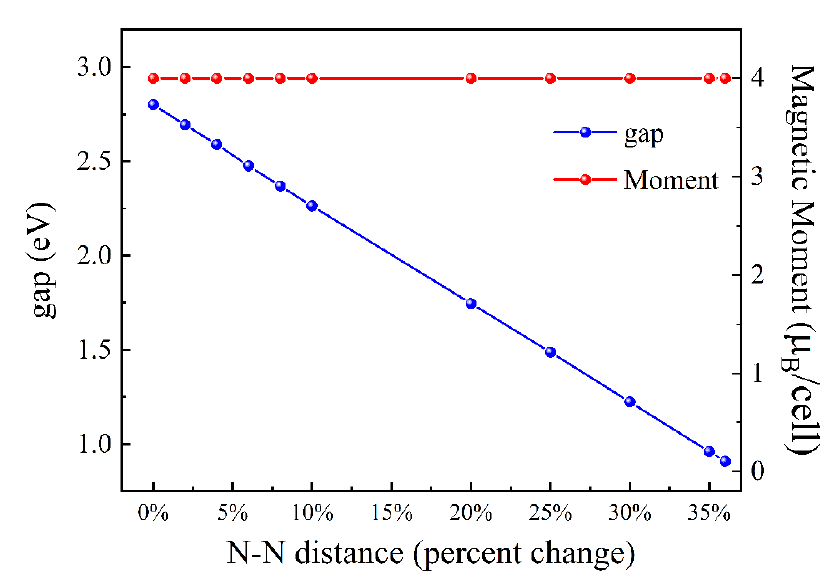}%
    	\caption{\label{fig6}The dependence of N-N distance on spin gap and magnetic moment.}
    \end{figure}

    The robustness of half metallicity is examined for FeN$_4$. Fig. \ref{fig5}(f) shows that the magnetic moment and energy difference ($\emph{E}$$_{FM}$-$\emph{E}$$_{AFM}$) under strains, where the strain was simulated by $\varepsilon$=(\emph{a}-\emph{a}$_0$)/\emph{a}$_0$$\times$100\%. The half metallicity keeps against AFM until a tensile strain $\sim$ 10\%, but collapses at a $\sim$ 4\% compressive strain. The large spin gap depends much on the splitting of $\pi$ and $\pi$$^*$ energy level. To check the effects of N-N bondlength on the spin gap of FeN$_4$, we enlarged the N-N distance while kept the Fe-N distance intact. As N-N being far away, $\pi$ and $\pi$$^*$ get close to each other, resulting in decreased spin gap but a constant magnetic moment, see Fig. \ref{fig6}. Hence, the large spin gap originates from the appropriate bondlength of N-N. These calculated results agree well with our proposition that N$_2$ dimers act as an electron receiver accommodating by their $\pi$$^*$ orbitals with a reduced electronegativity. The reduced electronegativity is consistent with the self-doping phenomenon in FeN$_4$. 
   
    \begin{figure}
	\includegraphics{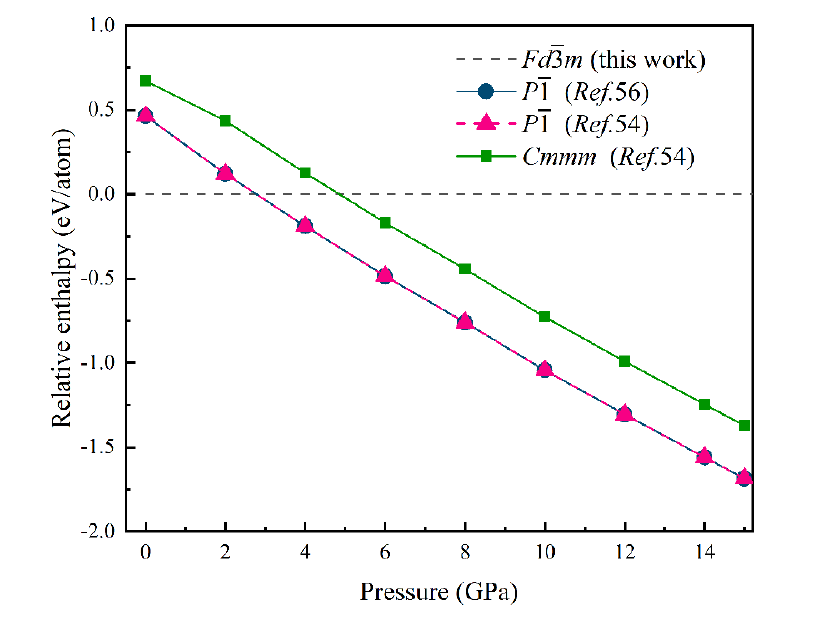}%
	\caption{\label{fig7}Relative enthalpies of FeN$_4$ with different structures as a function of pressure. Cubic (\emph{Fd$\bar{3}$m}) FeN$_4$ is taken as a standard. Negative values indicate other structures are more stable than cubic structure at a certain pressure.}
	\end{figure}

	Recently, $\emph{Chen et al.}$ \cite{RN52} predicts that FeN$_4$ has space groups of \emph{P$\bar{1}$} and \emph{Cmmm} under different pressures using CALYPSO methodology \cite{RN22}. $\emph{Bykov et al.}$ \cite{RN50} synthesized another FeN$_4$ with space group \emph{P$\bar{1}$} by high-pressure and laser-heating method. To compare the relative stability among them, we computed the enthalpies of them at different pressures, see Fig. \ref{fig7}. The results show that our diamond-like FeN$_4$ will transform to \emph{P$\bar{1}$} symmetry under 2.78 GPa. Hence, high pressures may not work for synthesizing FeN$_4$ we predict here. But it is similar to T-carbon \cite{RN34} in both structure and low density. The latter compound has been successfully synthesized by picosecond laser irradiation under a nitrogen atmosphere \cite{RN45}. It might be a choice to apply similar conditions to synthesize MN$_4$.

\section{Conclusions}	
	By first principles calculations, we predicted three half metals MnN$_4$, FeN$_4$ and CoN$_4$. They crystallize in a diamond-like structure with space group $\emph{Fd$\bar{3}$m}$. They are not only favored in energy, lattice-dynamically and mechanically stable, but also possess robust FM coupling with the highest Curie temperature \emph{T}$_c$ $\sim$ 10$^3$ K. Band structures indicate the largest spin gap is around 5 eV (HSE06). The proposition of self-doping of these 3\emph{d} transition metals caused by the reduction in electronegativity of N$_2$ dimers, well explains the calculated integer magnetic moments of MN$_4$. The N$_2$ dimer plays an important role in enlarging the spin gap for half metal MN$_4$ (M=Mn, Fe, Co). Other emergent magnetic properties are expected in compounds consisting of $\emph{d}$ metals and N$_2$ dimers with modified structures.
	
	\vspace{1 ex}
	
	This work is financially supported by the National Natural Science Foundation of China under Grants No. 51532010, and No. 51772322; the National Key Research and Development Program of China (2016YFA0300600, 2017YFA0304700); and the Key Research Program of Frontier Sciences, Chinese Academy of Sciences, Grant No. QYZDJ-SSW-SLH013.
	
%

\end{document}